\documentclass[pra,showpacs,preprint]{revtex4}
\usepackage{graphicx}
\usepackage[usenames]{color}
\usepackage{dcolumn}
\usepackage{amsmath}
\usepackage{amssymb}
\usepackage{bm}
\usepackage{rotating}
\usepackage{graphics}
\usepackage{epsfig}
\begin{document}
\input{epsf}
\preprint{APS/123-QED}
\title{Quenching of $para$-H$_2$ with an ultra-cold anti-hydrogen 
atom $\overline{\mbox{H}}_{1s}$}
\author{Renat A. Sultanov$^1$\footnote{rasultanov@stcloudstate.edu; 
r.sultanov2@yahoo.com}, Sadhan K. 
Adhikari$^2$\footnote{adhikari@ift.unesp.br, 
\; URL: http://www.ift.unesp.br/users/adhikari/}, and Dennis 
Guster$^1$\footnote{dcguster@stcloudstate.edu}}
\affiliation{$^1$High Performance Computing Laboratory, St. Cloud State 
University,
St. 
Cloud, MN 
56301-4498, USA\\
$^2$Institute of Theoretical Physics, UNESP
$-$ S\~ao Paulo State University,\\ 01140-070
S\~ao
Paulo, SP, Brazil}
\date{\today}    
\begin{abstract}

In this work we report the results concerning calculations for quantum-mechanical
rotational transitions in molecular hydrogen, H$_2$,  
induced by an ultra-cold ground state anti-hydrogen atom $\overline{\mbox{H}}_{1s}$.
The calculations are accomplished
using a non-reactive close-coupling quantum-mechanical approach.
The H$_2$ molecule is treated as a rigid rotor.
The total elastic scattering cross section $\sigma_{el}(\epsilon)$ 
at energy $\epsilon$,
state-resolved rotational transition cross sections $\sigma_{jj'}(\epsilon)$
between states $j$ and $j'$
and corresponding thermal rate coefficients $k_{jj'}(T)$
are computed in the temperature range 0.004 K $ \lesssim
T \lesssim$ 4 K.
Satisfactory agreement with other
calculations (variational) has been obtained for $\sigma_{el}(\epsilon)$.

\end{abstract}

\maketitle

\section{INTRODUCTION}

Interaction and collisional properties between matter and antimatter is of fundamental
importance in physics \cite{landua04,holzsch04}. The anti-hydrogen atom
$\overline {\mbox{H}}$, which is a bound state of an anti-proton $p^-$ and a 
positron $e^{+}$, is the simplest representative of an antimatter
atom.
This is a two-particle system, which, however, may  possess very
different interactional and dynamical properties compared to its matter 
counterpart the H atom \cite{voron05a,voron05b}.

By now much effort has been exerted in various experiments to build 
and store  $\overline {\mbox{H}}$ at cold and ultra-cold temperatures
\cite{collins05,amoretti02,hijmans02,gabri05,gabri}. New experiments are planned or in
progress to test the fundamental laws and theories of  physics 
involving antiparticles and antimatter in general \cite{holzsch04}.
For example, it follows from the CPT symmetry of quantum
electrodynamics
that a charged particle and 
its anti-particle counterpart should have equal and opposite charges and equal masses, 
lifetimes and gyromagnetic ratios. The CPT symmetry 
 predicts that hydrogen and anti-hydrogen atoms 
should have identical spectra. 
In this way experimentalists plan to test whether in fact H and $\overline {\mbox{H}}$ have
such  properties.
Specifically, a starting point would be to compare the frequency of the
1s-2s two-photon transition in H and $\overline {\mbox{H}}$.
Also, one of the 
important practical applications of antihydrogen has been mentioned in
\cite{nieto2004}, where the authors considered controlled $\overline {\mbox{H}}$
propulsion for NASA's future plans in very deep space.
Researchers at CERN \cite{landua04} and from other groups \cite{gabri05}
are interested 
 to trap and study $\overline {\mbox{H}}$ at  low temperatures, e.g., 
$T \lesssim 1$ K, 
when  the $\overline {\mbox{H}}$ atom will be  almost in its rest frame. 
The study of Lamb shift and response of antihydrogen to gravity at 
ultra-low energies should allow them to test more precisely
the predictions of two fundamental theories of modern physics: quantum
field theory and Einstein's general theory of relativity \cite{holz99}.

It has been pointed out that
the main cause of loss of $\overline {\mbox{H}}$ atoms confined  in a 
magnetic 
gradient trap is due to
$\overline {\mbox{H}}+$H$_2$ and $\overline {\mbox{H}}+$He 
collisions. Therefore, 
the $\overline {\mbox{H}}+$H$_2$ 
scattering cross-sections 
and corresponding rotational-vibrational 
thermal rate coefficients, in the case of H$_2$,
would be very helpful to gain a practical understanding of the slowing 
down and trapping of $\overline {\mbox{H}}$. 
Hence the investigation of the possibility of cooling of 
$\overline {\mbox{H}}$  atoms by
colliding them with colder H$_2$ 
is of significant 
practical interest \cite{cohen06}.
(Similar collision between trapped fermionic atoms with cold bosonic 
atoms has 
been fundamental in cooling the fermionic atoms and thus leading them 
to quantum degeneracy \cite{rmp}.)
Such investigation of $\overline {\mbox{H}}$ interaction with H and H$_2$
can reveal the survival conditions of $\overline 
{\mbox{H}}$ in
collisions with H and, even more importantly, with H$_2$ \cite{gregory08}.

Further, cooling occurs by energy transfer in
elastic collisions of $\overline {\mbox{H}}$ with H$_2$. However, during the 
collision the rearrangement process may
lead to the formation of protonium ($pp^-$) and positronium
($e^+e^-$) exotic atoms 
and the destruction of $\overline 
{\mbox{H}}$ atoms.
They are formed as matter-antimatter bound states, 
which
then annihilate. (There have also been many studies of scattering 
of positronium atoms \cite{ska}, the lightest matter-anti-matter atom).
Thus one can conclude,  that
the effectiveness of cooling $\overline {\mbox{H}}$
is determined from a comparison 
of the cross sections for direct scattering and rearrangement.

By now a series of theoretical works have been published, in which the 
properties of interaction
between $\overline {\mbox{H}}$ and H, He, H$_2$ have been  investigated 
\cite{stras02,stras02b,stras04,stras05}.
Some theoretical studies have been carried out for the $\overline 
{\mbox{H}}$+H
system at thermal energies using quantum-mechanical methods 
\cite{armour99,ghosh00,armour00,froel01,armour02,ghosh03}.
Also, discussions on the importance and applications for this system, 
especially,
in connection with Bose-Einstein condensation \cite{nieto},
ultra-cold collisions \cite{ghosh00,gregory08}, and its static and dynamic properties 
\cite{other},
can be found in the literature.

In this work we present results for the collision of an ultra-cold
$\overline {\mbox{H}}$ atom 
 with  H$_2$,  where
H$_2$ is treated as a rigid rotor with a fixed distance between
hydrogen atoms.
The elastic, rotational state-resolved scattering cross sections 
for the $\overline {\mbox{H}}$-H$_2$ scattering 
and their corresponding thermal rate coefficients are calculated using a 
non-reactive 
quantum-mechanical close-coupling approach. The potential interaction
between $\overline {\mbox{H}}$ and the hydrogen atoms is taken from Ref.  
\cite{stras02}.

In the next section we  present the quantum-mechanical formalism  used in 
this work. The results and discussion are presented in 
Sec. III. Conclusions are given in Sec. IV.

\section{$\overline {\mbox{H}}$-H$_2$ Scattering Formulation}

\subsection{Basic Equations}

In this section we describe  the close-coupling quantum-mechanical 
approach we used to calculate
the cross sections and 
collision rates of a hydrogen molecule H$_2$
with an anti-hydrogen atom $\overline {\mbox{H}}$.
Atomic units 
($e=m_e=\hbar=1$) are used in this section, 
where $e$ and $m_e$ are charge and mass of an electron.
Three-body Jacobi coordinates $\{\vec r,\vec R\}$ for the
$\overline {\mbox{H}}$+H$_2(j)$ system used in this work are shown in Fig.\ 
1. The two H atoms are labeled 2 and 3 and the $\overline{\mbox{H}}$ atom is
labeled 1, $O$ is the center of mass of the H$_2$ molecule,
$\Theta$ is the polar angle between vector $\vec r$ connecting the two H 
atoms 
in H$_2$ (labeled  2 and 3) and vector $\vec R$ connecting the center of 
mass of the H$_2$ molecule to the  $\overline{\mbox{H}}$ atom (labeled
1). Next,
$\vec j$ and $\vec L$ are angular momenta corresponding to the vectors 
$\vec r$ and $\vec R$,  respectively.  
The quantities 
 $x_{21}$ and $x_{31}$ are  the distances between the $\overline{\mbox{H}}$ 
atom labeled 1 and the H atoms labeled 2 and 3, respectively.

The Schr\"odinger equation for an $a+bc$ collision in the center of 
mass frame, where $a$ $(\overline{\mbox{H}})$ is an atom and 
$bc$ $(\overline{\mbox{H}}_2)$ is a linear rigid rotor, is
\cite{arthurs,green79}
\begin{equation}
\left(\frac{P_{\vec R}{^2}}{2{M_R}} + \frac{L_{\hat 
r}{^2}}{2\mu r^2} + V(\vec r,\vec R) - E \right)
\Psi(\hat r,\vec R)=0.
\label{eq:schred}
\end{equation}
where $P_{\vec R}$ is the relative momentum between $a$ and $bc$,  
{${M}_R$} is the  reduced
mass of the atom-molecule (rigid rotor in this model) system $a+bc$:
${M_R} = m_a(m_b+m_c)/(m_a+m_b+m_c)$,
$\mu$ is the  reduced mass of the target:
$\mu=m_am_b/(m_a + m_b)$,
$\hat r$ is the angle of orientation of the rotor $ab$, 
$V(\vec r, \vec R)$ is the potential energy surface (PES) for the 
three-atom system $abc$, and
$E$ is the total energy of the system.
The eigenfunctions of the operator $L_{\hat r}{^2}$ 
in Eq. (\ref{eq:schred}) are the spherical harmonics $Y_{jm}(\hat r)$.

To solve Eq.  (\ref{eq:schred}), the following expansion is used \cite{green75}
\begin{equation}
\Psi(\hat r,\vec R)=\sum_{JMjL}\frac{U^{JM}_{jL}(R)}{R}
\phi^{JM}_{jL}(\hat r,\vec R),
\label{eq:expan}
\end{equation}
where channel expansion functions are
\begin{eqnarray}
\phi^{JM}_{jL}(\hat r,\vec R) = \sum_{m_1m_2}
C_{jm_1Lm_2}^{JM}  Y_{jm_1}(\hat r)  Y_{Lm_2}(\hat R),
\end{eqnarray}
here $\vec J = \vec j + \vec L$ is the total angular momentum of the system
$abc$,
and 
$M$ is its
projection onto the space fixed $z$ axis, 
$m_1$ and $m_2$ are projections of $j$ and $L$ respectively, 
$C_{jm_1Lm_2}^{JM}$ are the Clebsch-Gordan coefficients, and $U$'s are 
the appropriate radial functions.

Substitution of Eq. (\ref{eq:expan}) into Eq. (\ref{eq:schred}) 
provides a set of coupled second order 
differential equations
for the unknown radial functions $U^{JM}_{jL}(R)$
\begin{eqnarray}
&&\left(\frac{d^2}{dR^2}-\frac{L(L+1)}{R^2}+k_{jL}^2\right)U_{jL}^{JM}(R)
=2 {M_R}
\sum_{j'L'} \int <\phi^{JM}_{jL}(\hat r,\vec R)  
|V(\vec r,\vec R)| 
\phi^{JM}_{j'L'}(\hat r,\vec R)> \nonumber \\
&&\times U_{j'L'}^{JM}(R) d\hat r d\hat R.
\label{eq:cpld}
\end{eqnarray}
To solve the coupled radial  equations (\ref{eq:cpld}), we 
apply the hybrid modified log-derivative-Airy propagator in the
 general purpose scattering program MOLSCAT 
\cite{hutson94}. 
Additionally, we tested other
propagator schemes included in MOLSCAT.
Our calculations reveal that other propagators can also produce 
quite stable results.


The log-derivative matrix is propagated to large intermolecular 
distances $R$, since all 
experimentally observable
quantum information about the collision is contained in the asymptotic 
behavior of functions 
$U^{JM}_{jL}(R\rightarrow\infty)$. 
The numerical results are matched to the known asymptotic behavior 
of $U^{JM}_{jL}(R)$ relating to the the physical scattering $S$-matrix 
\cite{landau} 

\begin{eqnarray}
U_{jL}^{JM}
&\mathop{\mbox{\large$\sim$}}\limits_{R \rightarrow + \infty}&
\delta_{j j'} \delta_{L L'}
e^{-i(k_{\alpha}R-(L\pi/2))} 
 - \left(\frac{k_{\alpha}}{k_{\alpha'}}\right)^{1/2}S^J(j'L';jL;E) 
e^{i(k_{\alpha'}R-(L'\pi/2))},
\end{eqnarray}
where $k_{\alpha}=[2{M_R}(E-E_{\alpha})]^{1/2}$ is the 
channel wave-number of channel $\alpha=(jL)$,
$E_{\alpha}$ is rotational channel energy and $E$ is the total energy in the 
$abc$ system.
This method was used for each partial wave until a converged cross section was obtained. 
It was verified that the results have converged with respect to the 
number of partial waves as well as
the matching radius, $R_{max}$, for all channels included in our calculations.

Cross sections for rotational excitation and relaxation phenomena can be 
obtained directly from the $S$-matrix.
In particular, the cross sections for excitation from $j\rightarrow j'$ 
summed over the final $m'$ and averaged over the initial $m$ are given by
\cite{green75}
\begin{eqnarray}
\sigma(j',j,\epsilon)&=&\frac{\pi}{(2j+1)k^2_{\alpha}}
\sum_{JLL'}(2J+1)|\delta_{jj'}\delta_{LL'} -            
S^J(j'L';jL; E)|^2.
\label{eq:cross}
\end{eqnarray}
The kinetic energy is 
$\epsilon=E-B_ej(j+1)$,
where $B_e$ is the rotation constant of the rigid rotor $bc$, i.e. the hydrogen molecule.

The relationship between the rate coefficient $k_{j\rightarrow j'}(T)$ and the corresponding
cross section $\sigma_{j\rightarrow j'}(E_{kin})$ can be obtained through the following
weighted average 
\cite{gert}
\begin{equation}
k_{j\rightarrow j'}(T) = \frac{8k_BT}{\pi{M_R}}\frac{1}{(k_BT)^2}\int_{\epsilon_s}^{\infty}
\sigma_{j\rightarrow j'}(\epsilon)e^{-\epsilon/k_BT}\epsilon d\epsilon,
\end{equation}
where $\epsilon = E - E_{j}$ is pre-collisional translational energy at 
temperature $T$, $k_B$ is Boltzman constant, and $\epsilon_s$ is the minimum value of the
kinetic energy needed to make $E_j$ levels accessible.

\subsection{$\overline {\mbox{H}}-$H$_2$ Interaction Potential}

In the following section, we will present our results for rotational 
quantum transitions in
collision between H$_2$ and an anti-hydrogen atom
$\overline{\mbox{H}}$, that is
\begin{equation}
\mbox{H}_2(j) + \overline{\mbox{H}} \rightarrow \overline{\mbox{H}} + \mbox{H}_2(j').
\label{eq:proc1}
\end{equation}
Here H$_2$ is treated as a vibrationally averaged rigid monomer rotor.
The bond length was fixed at 1.449 a.u. or 0.7668 \AA. The rotation constant of
the H$_2$ molecule has been taken as $B_e=60.8$cm$^{-1}$.
The H$_2$ rigid rotor model has been already applied in different
publications 
\cite{green77,schaefer1990,flower98,roueff99,renat06,renat0607,renat09,green75}.
For the considered range of kinetic energies
the model can be quite justified in this special case when
only pure rotational quantum transitions at low collisional energies are of
interest as in H$_2$($j$)+$\overline{\mbox{H}}$,
and when the energy gap between rotational and vibrational energies is
much larger  than kinetic energy of collision.
In such a model the quantum
mechanical approach is rather simplified.

Next we 
consider an important physical parameter in
atomic and molecular collisions, e. g., the PES  between
the atoms.
There is no
global potential energy surface available  for the
three-atom $\overline{\mbox{H}}$-H$_2$ system. However in Ref. \cite{stras02},
the author  calculated the values of interaction energy between H
and $\overline{\mbox{H}}$, i. e.,  the H-$\overline{\mbox{H}}$ energy
curve using the Rayleigh-Ritz variational method.
Further, the microHartree accuracy of Born-Oppenheimer energies of the system
has been achieved in that work.

To construct the H$_2$-$\overline{\mbox{H}}$ interaction potential we take
the H-$\overline{\mbox{H}}$
energy data from Ref. \cite{stras02} and make a cubic spline
interpolation through all 46 points taken from Table 1 of that paper.
These data have been tabulated from $R_{min}=0.744$ a.u. to $R_{max}=20.0$
a.u. inter-atomic distances. Because in the current work we use the rigid rotor
model for H$_2$, we do not need 
the interaction energy
between hydrogen atoms in H$_2$.
The interaction potential between a hydrogen molecule and
$\overline{\mbox{H}}$ is taken by sandwiching the two $\overline{\mbox{H}}$-H
potential energy curves:
\begin{equation}
V(\vec r, \vec R) = V(r, R, \Theta) = V^{21}_{\text{H}-\overline {\text{H}}}(x_{21})
+ V^{31}_{\text{H}-\overline {\text{H}}}(x_{31}),
\label{eq:pesmodel}
\end{equation}
where distances between atoms are written as follow (cf. Fig. 1):
\begin{eqnarray}
x_{21} = \sqrt{r^2/4 + R^2 + rR 
\cos \Theta} 
\hspace{6 mm} \text{and} \hspace{6 mm}  
x_{31} = \sqrt{r^2/4 + R^2 + rR 
\cos (\pi - \Theta)}.
\end{eqnarray}

The functions  $V^{k1}_{\overline {\text{H}}-\text{H}}(y)$ with $k=2(3)$
are represented as cubic spline interpolation functions for any value of $y=x_{21}$ or
$y=x_{31}$ as follows:
\begin{equation}
V^{k1}_{\overline {\text{H}}-\text{H}}(y) = V^{k1}_{\overline
{\text{H}}-\text{H}}(X_i) + B_i(y - X_i)
+ C_i(y - X_i) + D_i(y - X_i),
\end{equation}
where $B_i(y - X_i)$, $C_i(y - X_i)$, $D_i(y - X_i)$ perform the
spline interpolation and where 
$X_i \leqslant y \leqslant X_{i+1}$, in each sub-interval $[X_i,
X_{i+1}]$, $i=1,2,3,...,(n-1), n=46$. The coordinates $X_i$ and
corresponding values of the H-$\overline{\mbox H}$ potential energy data
have been taken from Table 1 of Ref. \cite{stras02}.

The calculated PES is shown in Fig.\ 2. 
It is clear 
that the potential has a
singular value when the distance between  $\overline{\mbox{H}}$
and H$_2$ is equal to zero.
Additionally, the $\overline {\text{H}}$-H$_2$ 
PES which we obtain
from Eq. (\ref{eq:pesmodel}) is shown in Fig.\ 3. Specifically, this
potential was used in our calculations of $\overline{\mbox{H}}$ + H$_2$
collisions.
Again, as seen in Fig.\ 2
the potential energy curve between $\overline{\mbox{H}}$-H 
has a Coulomb type singularity at small distances.
In our calculations we needed to make
additional test runs to achieve convergence in our results.
In the next section we will briefly demonstrate the numerical
convergence of the results when calculating
total elastic scattering cross sections.
These results depend on various numerical and quantum-mechanical 
scattering parameters.

\section{RESULT and  DISCUSSION}

\subsection{Convergence test}

Numerous test calculations have been undertaken to insure the
convergence of the results with respect to all parameters that enter
in the propagation of the Schr\"odinger equation. These include the
atomic-molecular distance $R$, the total angular momentum $J$,
the number of total rotational levels to be included in the close-coupling
expansion and others, see Fig.\ 1.
Particular attention has been given to the total number of numerical
steps in the propagation over the distance $R$ of the Schr\"odinger equation
(\ref{eq:cpld}). Specifically, the parameter R ranges from 0.75 a.u. to 20.0 a.u.
We used up to 50000 propagation points.
We also applied and tested different mathematical propagation schemes included in MOLSCAT.

The rotational energy levels of $para$-H$_2(j)$ and the corresponding
angular momenta  $j$ are shown in Table I. The goal of this
work is to get new results for $\overline {\mbox{H}}$ + $para$-H$_2$
thermal rate coefficients $k_{j\rightarrow j'}(T)$ at
ultra-low temperatures: specifically $0.004$ K$< T < 4$ K. The
corresponding cross sections have 
been calculated for collision energies varying  from $\sim 0.0001$
cm$^{-1}$ to $\sim 100$ cm$^{-1}$. 
These energy values are very small.
However, despite this fact, to reach convergence of the results
we needed to include in  expansion (\ref{eq:expan})
a significant number of rotational levels of the H$_2$
molecule, specifically up to $j_{max}\approx 60$.
Below in Table II we present these results.

Also, we found that, at lower energies, for the numerical solution of
Eq.  (4)
a  much larger number of propagation (integration) points 
are needed than at higher energies. Specifically, at higher energies
we need 500 propagation points, but for lower energies 50,000 points are 
needed to achieve comparable precision.
Convergence has been achieved for
elastic scattering cross sections for various scattering parameters.
Below in Table III we present these results.
Then we used this data in our 
calculation for rotational energy transfer,
elastic scattering cross section and thermal rate coefficient.

Now in Table II we present results for the total elastic cross
sections for two collisional energies: 0.1 cm$^{-1}$ and 0.01 cm$^{-1}$
The cross sections are shown for a number of different maximum values of
the rotational angular momentum $j=j_{max}$ in the H$_2$ molecule
included in expansion (\ref{eq:expan}).
This is also the JMAX parameter in MOLSCAT \cite{hutson94}. 
Other scattering parameters have also been treated correctly in the
calculation. One can see that JMAX should be at least 56. The other
scattering parameter in Table II  is MXSYM \cite{hutson94}. It reflects the number
of terms in the potential expansion over angular functions \cite{green79,green75}.
It is therefore evident that in this
calculation 
we need to keep at least 24 terms in the expansion.

In Table III we also present results for total elastic scattering
cross sections for few more selected energies. However in this table
the convergence has been reached by increasing the total angular momentum 
$J$ 
and by increasing the total number of the propagation steps 
in the
propagation over the coordinate $R$ of the Sch\"odinger equation (\ref{eq:cpld}).
As expected, for lower energy collision we needed smaller values for
the maximum $J$. For example, for collision energy $E_{coll}=0.01$ a.u. 
it
is enough to have $J=0$, however $J=10$ should be taken for
$E_{coll}=100.0$ a.u.

In regard to the
total number of the propagation steps in the solution of Eq. 
(\ref{eq:cpld}) 
one can see, that we need to include many more propagation points 
for low energy calculations.
All test calculations in Table III have been done with JMAX = 56 and MXSYM = 24.
The obtained results concerning the numerical and scattering parameters
have been used in our  
calculation for total elastic $\sigma_{el}(E)$
and rotational quantum state transfer $\sigma_{j\rightarrow j'}(E)$
cross sections and corresponding thermal rate coefficients $k_{j\rightarrow j'}(T)$.


\subsection{$\overline {\mbox{H}}$ + $para$-H$_2$ results}


Now we 
present computational results for  process (\ref{eq:proc1}), namely, for
elastic scattering ($j=0\rightarrow j'=0$) and for low quantum number 
rotational transition
between levels with $j=0, 2$ and 4:
$2\rightarrow 0$, $0\rightarrow 2$, $4\rightarrow 0$, and $4\rightarrow 2$.
From the results of Table III
we see that to reach numerical convergence, for example, for the
elastic scattering cross section, we need to include a large number of
H$_2$ rotational levels, specifically up to 60.

The results for the elastic scattering cross sections $\sigma_{el}(\epsilon)$ for
$\overline {\mbox{H}}$ + H$_2$ $\rightarrow$ H$_2$ + $\overline {\mbox{H}}$
are shown in Fig.\ 4 together with the corresponding results of variational
calculations of Gregory and Armour \cite{gregory08}. 
It can be seen, that basically
the two sets of  cross sections are
close to each other, although in our calculation we use larger number
of collision energy points, specifically up to 200.
In our calculation a shape resonance is found at energy
$\epsilon \sim 3.5 \times 10^{-5}$ Hartree. As in Ref. \cite{gregory08}
our $\sigma_{el}(\epsilon)$ tends to reach a 
constant value at lower energies with  $\sigma_{el}(\epsilon \lesssim
10^{-8}\  a.u. )=9.47 \times 10^3 a_0^2$.
This result allows us to calculate the $\overline{\mbox{H}}$+H$_2$
scattering length, which is
\begin{equation}
a = \sqrt{\sigma_{el}/(4\pi)} = 27.5\ a_0. 
\end{equation}
The Gregory-Armour  scattering length \cite{gregory08} obtained with a variational method is
$\tilde a=19.5\ a_0$.  
The two results are in reasonable agreement with each other.

In Fig. \ 5 (a) and (b) we show the total state-resolved cross sections 
$\sigma_{j=2\rightarrow j'=0}$(v) vs. velocity v and the corresponding 
thermal rate coefficients $k_{2\rightarrow 0}$(T) vs. temperature T for 
the hydrogen molecule rotational relaxation process. It is seen, that 
when the collision energy increases the de-excitation cross section 
decreases. It can be explained in the following way: at low relative 
velocities (kinetic energies) between H$_2(j=2)$ and 
$\overline{\mbox{H}}$, the rotationally excited H$_2$ molecule has more 
time for interaction and consequently, it has higher quantum-mechanical 
probability to release its internal rotational energy to 
$\overline{\mbox{H}}$. The resulting corresponding rate coefficients 
have been calculated for a temperature range from $0.004\ K <\ T\ < 4\
K$ and are also presented in Fig.\ 5 below the cross section results.

Next, in Fig.\ 6 (a) and (b) we present  results for  
the total state-resolved cross sections
$\sigma_{0\rightarrow j'=2}$(v) vs. velocity v and the corresponding
thermal rate coefficients $k_{0\rightarrow 2}$(T) vs. temperature T for
the hydrogen molecule rotational excitation process.
It is quite understandable, 
as we find from Fig. 6 (a),  that
when the collision energy (relative velocity v)
is increases the quantum-mechanical probability and corresponding
cross section of the rotational excitation of H$_2(j)$
are also increases. Figure 6 (b) depicts the corresponding results 
for the thermal rate coefficient.

In Figs. 7 (a) and (b) we present 
results for cross sections and rates for rotational 
relaxation process, as in Figs. 5 (a) and (b), but now connecting the 
states $j=4$ and $j'=2$. 
Finally, in Figs. 8 (a) and (b) we present 
results for cross sections and rates 
for rotational 
relaxation process but now connecting the 
states $j=4$ and $j'=0$. 
An unexpected result has been found in Fig. 7 (a) 
in the rotational transition
cross section $\sigma_{4\rightarrow 2}(v)$, 
i.e., when the  H$_2$ quantum angular momentum
has been changed from $j=4$ to $j'=2$. 
One can see, that the values of these cross sections at very low
collision energies are almost from 5 to 10 times larger then other
cross sections considered in this work, compare with the results
from Figs. 5, 6 and 8.

\section{SUMMARY}

A quantum-mechanical study of the state-resolved rotational relaxation and
excitation cross sections and thermal rate coefficients in ultra-cold collisions
between hydrogen molecules H$_2$ and anti-hydrogen atoms $\overline{\mbox{H}}$
has been the subject of this work. 
A model PES for H$_2$-$\overline{\mbox{H}}$
has been constructed by sandwiching two H-$\overline{\mbox{H}}$
interaction potentials for  two different hydrogen atoms taken from  Ref. \cite{stras02}.
This H-$\overline{\mbox{H}}$ interaction potential 
is  shown in Fig.\ 2. 
The   H$_2$-$\overline{\mbox{H}}$ PES is presented in Fig.\ 3.
Calculation for total elastic scattering cross section
and for low quantum rotational transition states have been performed.
We considered only the following quantum transitions:
2$\rightarrow$0, 0$\rightarrow$2, 4$\rightarrow$2, and
4$\rightarrow$0.

A test of the numerical convergence was 
undertaken. These results are presented in Tables II and III.
%
%
Our results reveal that it is necessary to set 
the rotational angular momentum $j_{max}$ in the H$_2$ molecule
to a relatively large  number, i.e. in the expansion (\ref{eq:expan})
we needed to include up to 60 terms.
The calculation was performed using the  MOLSCAT program \cite{hutson94}.
Different propagation schemes included in the MOLSCAT program  have
 been used and
tested. Additionally, the MXSYM potential parameter in that program 
also needed to have a
relatively large  value to obtain good convergence 
as can be seen from  Table II. 
The numerical convergence
has also been tested over the total number of the propagation steps
over coordinate $R$
in the solution of the Schr\"odinger equation (\ref{eq:cpld}). We have
found, that at low energies we need a much larger  number of 
integration points than at higher  energies, cf. Table III.
Our results for the H$_2(j)$+$\overline{\mbox{H}}$ total elastic scattering
cross section are in reasonable agreement with the corresponding results from
Gregory and Armour \cite{gregory08}. The authors of this paper used a different PES,
which is still unpublished, and
applied a quantum-mechanical variational approach.
Unfortunately, the rotational transitions in the
H$_2(j)$+$\overline{\mbox{H}}$ collisions have not been
calculated in that work \cite{gregory08}.
One of the interesting results of the present  work is that the cross
section of the rotational transitions from H$_2(j=4\rightarrow j=2)$
at ultra-low energies are approximately 5-10 times larger than other
transition state cross sections.

To the  best of our knowledge we do not know of
any other calculation of the
rotational transitions in
the H$_2(j)$+$\overline{\mbox{H}}$ collision. 
 These results can help to model energy transfer 
processes in the hydrogen-anti-hydrogen plasma, 
and perhaps to design new experiments
in the field of the anti-hydrogen physics.
Finally, we believe, that in the future work
it should be useful to include vibrational degrees 
of freedom of the
H$_2$ molecules, i.e. to carry out 
 quantum-mechanical
calculations for different rotational-vibrational relaxation processes:
$
\mbox{H}_2(v,j) + \overline{\mbox{H}} \rightarrow \overline{\mbox{H}} + \mbox{H}_2(v',j'),
$
where $v$ and $v'$ are the vibrational quantum numbers of H$_2$
before and after the collision, respectively.

\vspace{5mm}

\noindent{{\bf ACKNOWLEDGMENT}}

This work was supported by St. Cloud State University internal grant program and 
CNPq and FAPESP of Brazil.

\clearpage

\begin{table*}
\caption{$para$-H$_2$ rotational spectrum.}
%
\label{table:1}
\begin{ruledtabular}
\begin{tabular}{ccc}
 Level &  Rotational energy (cm$^{-1}$) & Internal quantum momentum in $para$-H$_2$(j)\\
\hline
   1   &        0.00 &    0\\
   2   &      364.80 &    2\\
   3   &     1216.00 &    4\\
   4   &     2553.60 &    6\\
   5   &     4377.60 &    8\\
   6   &     6688.00 &   10\\
   7   &     9484.80 &   12\\
   8   &    12768.00 &   14\\
   9   &    16537.60 &   16\\
  10   &    20793.60 &   18\\
  11   &    25536.00 &   20\\
  12   &    30764.80 &   22\\
  13   &    36480.00 &   24\\
  14   &    42681.60 &   26\\
  15   &    49369.60 &   28\\
  16   &    56544.00 &   30\\
  17   &    64204.80 &   32\\
  18   &    72352.00 &   34\\
  19   &    80985.60 &   36\\
  20   &    90105.60 &   38\\
  21   &    99712.00 &   40\\
  22   &   109804.80 &   42\\
  23   &   120384.00 &   44\\
  24   &   131449.60 &   46\\
  25   &   143001.60 &   48\\
  26   &   155040.00 &   50\\
  27   &   167564.80 &   52\\
  28   &   180576.00 &   54\\
  29   &   194073.60 &   56
\end{tabular}
\end{ruledtabular}
\end{table*}

\clearpage

\begin{table*}
\caption{Convergence of the total elastic scattering 
cross section
$\sigma_{el}$ $(10^{-16}\mbox{cm}^2)$ 
at different energies $E$ (cm$^{-1}$)
in $\overline{\text{H}}$+H$_2\rightarrow$ H$_2$+$\overline{\text{H}}$
with respect to the maximum value of the rotational angular
momentum $j=j_{max}$ in H$_2(j)$ included in the expansion
(\ref{eq:expan}) (parameter JMAX in MOLSCAT). Convergence
with 
 the number of terms in the potential expansion (parameter MXSYM in MOLSCAT)
is also shown. Numbers in parentheses are powers of 10.
}
\centering
\label{table:2}
\begin{ruledtabular}
\begin{tabular}{lccccccccccc}
$E$ (cm$^{-1}$) & \multicolumn{10}{c} {$\sigma_{el} \times$ 10$^{16}$ (cm$^2$)}\\

\hline
& \multicolumn{5}{c}{JMAX} & \multicolumn{5}{c}{MXSYM}\\
\hline
& 30 & 40 & 50 & 56 & 60 & & 12 & 20 & 24 & 26 \\ 
\hline
0.1 & 61.0&5.25(2)&1.59(3)&1.59(3)&1.59(3)&  &1.97(3)&1.61(3)&1.59(3)&1.59(3)\\
0.01& 55.4&1.06(3)&6.59(3)&6.62(3)&6.62(3)&  &1.12(4)&6.75(3)&6.56(3)&6.54(3)
%
%
\end{tabular}
\end{ruledtabular}
\end{table*}

\clearpage

\begin{table*}
\caption{Convergence for the total elastic scattering cross section 
$\sigma_{el}$
$(10^{-16}\mbox{cm}^2)$ at different collision energies $E$ in (cm$^{-1}$)
in $\overline{\text{H}}$+H$_2\rightarrow$ H$_2$+$\overline{\text{H}}$
with respect to the maximum value of the total angular
momentum $J$  
of the 3-atomic system:
parameter JTOT in MOLSCAT. Convergence
on the number of numerical 
space steps  in propagation over distance $R$   of the
Schr\"odinger equation (parameter STEPS in MOLSCAT) is also shown.
Numbers in parentheses are powers of 10. }
\centering
\label{table:3}
\begin{ruledtabular}
\begin{tabular}{lcccccccc}
$E$ (cm$^{-1}$) & \multicolumn{8}{c}
{$\sigma_{el}  \times
10^{16}$ (cm$^2$)}
\\
\hline
&\multicolumn{3}{c}{JTOT} & \multicolumn{5}{c}{STEPS}\\
\hline
& 0 & 2 & - & & 500 & 1000 & 10000 & 50000 \\
\hline
0.1 & 1.59(3)&1.59(3)&       &   &3.64(2)&2.67    &1.60(3)&1.59(3)\\
0.01& 6.62(3)&6.62(3)&       &   &3.62(2)&1.75(-1)&6.63(3)&6.62(3)\\
\hline
\hline\\
& 4 & 6 & 8 & & 500 & 1000 & 5000 & 7000 \\
\hline
10.0  & 5.85(1)&5.96(1)&5.96(1)&   &6.23(1)&6.00(1)&5.96(1)&5.96(1)\\
1.0   & 1.69(2)&1.69(2)& - &       &1.77(2)&1.70(2)&1.70(2)&1.70(2)\\
\hline
\hline\\
& 8 & 10 & 12 & & 500 & 750 & - & - \\
100.0 & 1.65(2)&1.72(2)&1.72(2)&   &1.72(2)&1.72(2)& - & -
%
%
\end{tabular}
\end{ruledtabular}
\end{table*}

\clearpage


\begin{figure}
\begin{center}
\includegraphics[width=\linewidth]{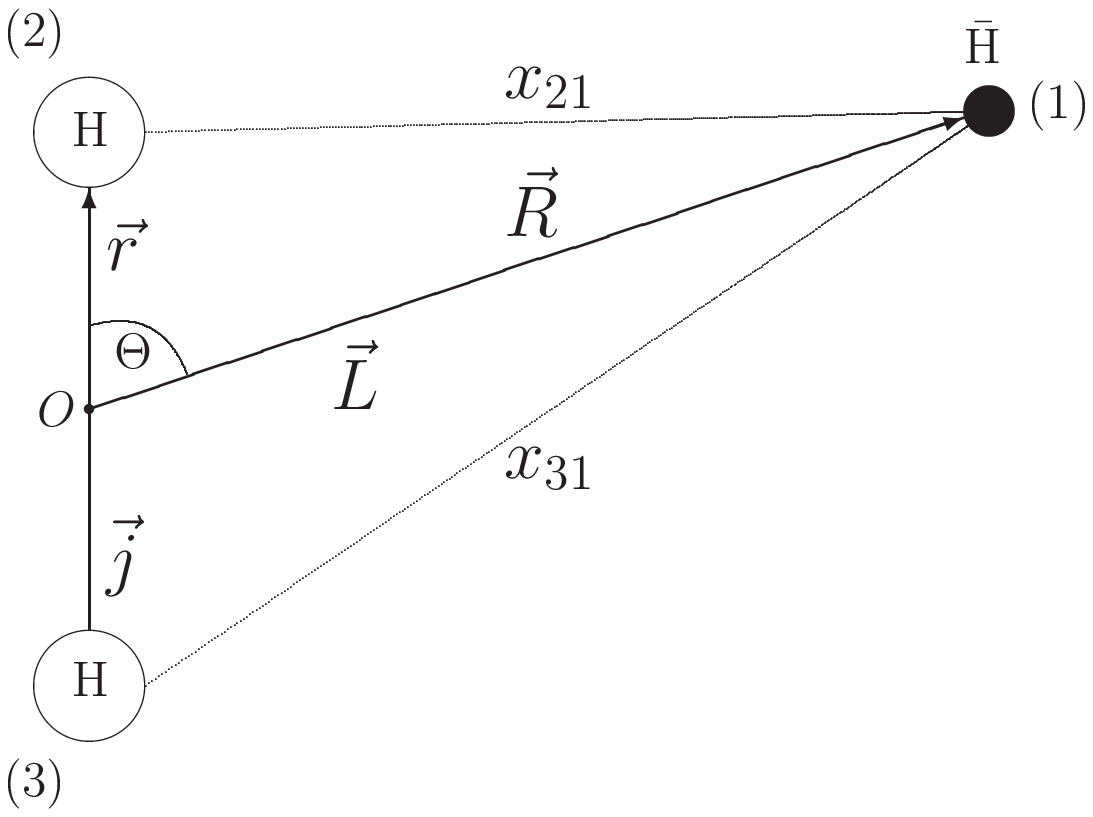}
\end{center}
\caption{Three-body Jacobi coordinates $\{\vec r,\vec 
R\}$ for the
$\overline {\mbox{H}}$+H$_2(j)$ system used in this work.
}\label{fig1}
\end{figure}

\clearpage

\begin{figure}
\begin{center}
\includegraphics*[scale=1.0,width=27pc,height=16pc]{HAH.eps}
\end{center}
\caption{$\overline {\mbox{H}}$-H potential energy
curve   from Ref. \cite{stras02}}
\label{fig3}
\end{figure}

\clearpage

\begin{figure}
\begin{center}
\vspace{-4.0 cm}
\includegraphics[width=\linewidth]{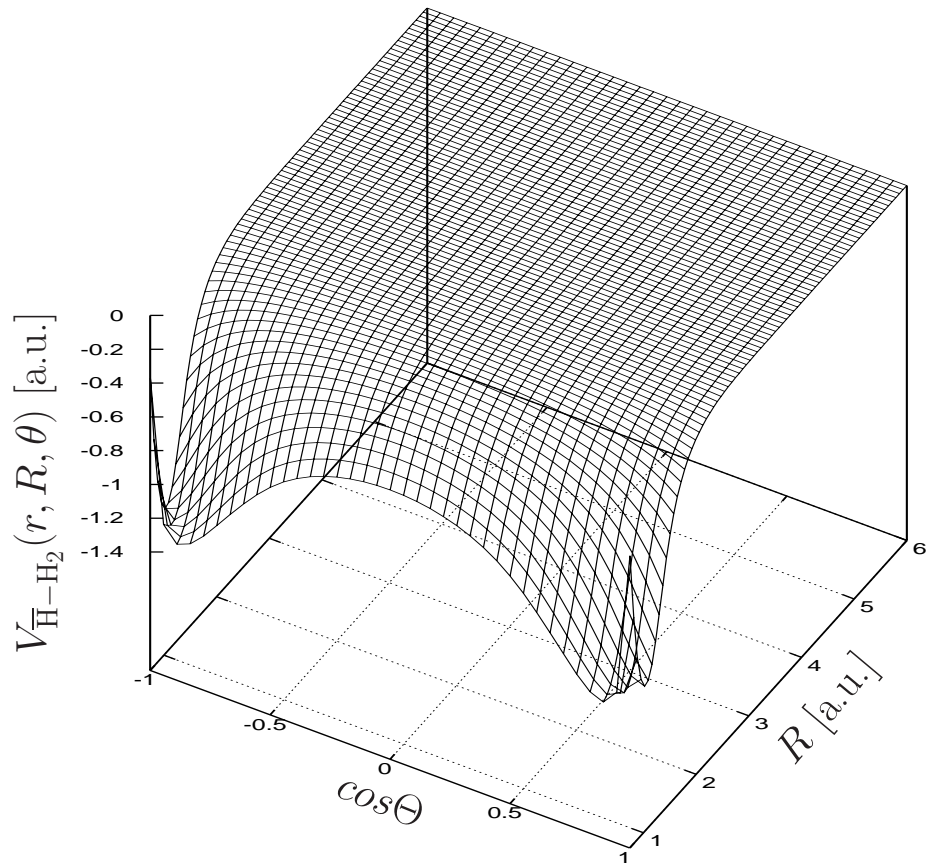}
\end{center}
\vspace{-4.0 cm}
\caption{
Interaction potential 
$V_{\overline{\text{H}}-\text{H}_2}(r,R,\Theta)$
between $\overline{\text{H}}$ and H$_2$  in a.u.
The distance between hydrogen atoms in H$_2$ is fixed at 
$r=r(\text{H}_2)=1.409$ a.u.}
\label{fig4}
\end{figure}

\clearpage

\begin{figure}

\begin{center}
\includegraphics*[scale=1.0,width=27pc,height=16pc]{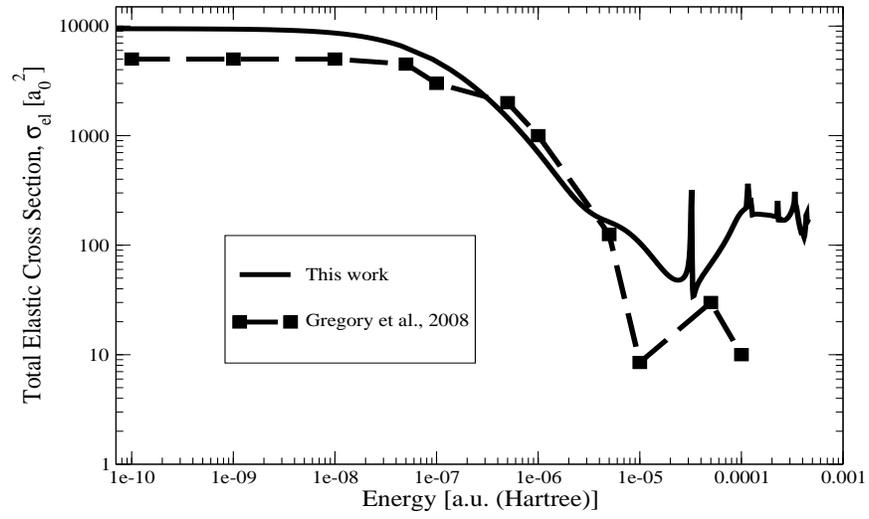}
\end{center}

\caption{Total elastic scattering cross section for 
$\overline {\mbox{H}}$ +
H$_2$ at different energies: results from Gregory {\it et al.} \cite{gregory08} and this work.
}\label{fig2}
\end{figure}

\clearpage

\begin{figure}
\begin{center}
\includegraphics*[scale=1.0,width=27pc,height=15pc]{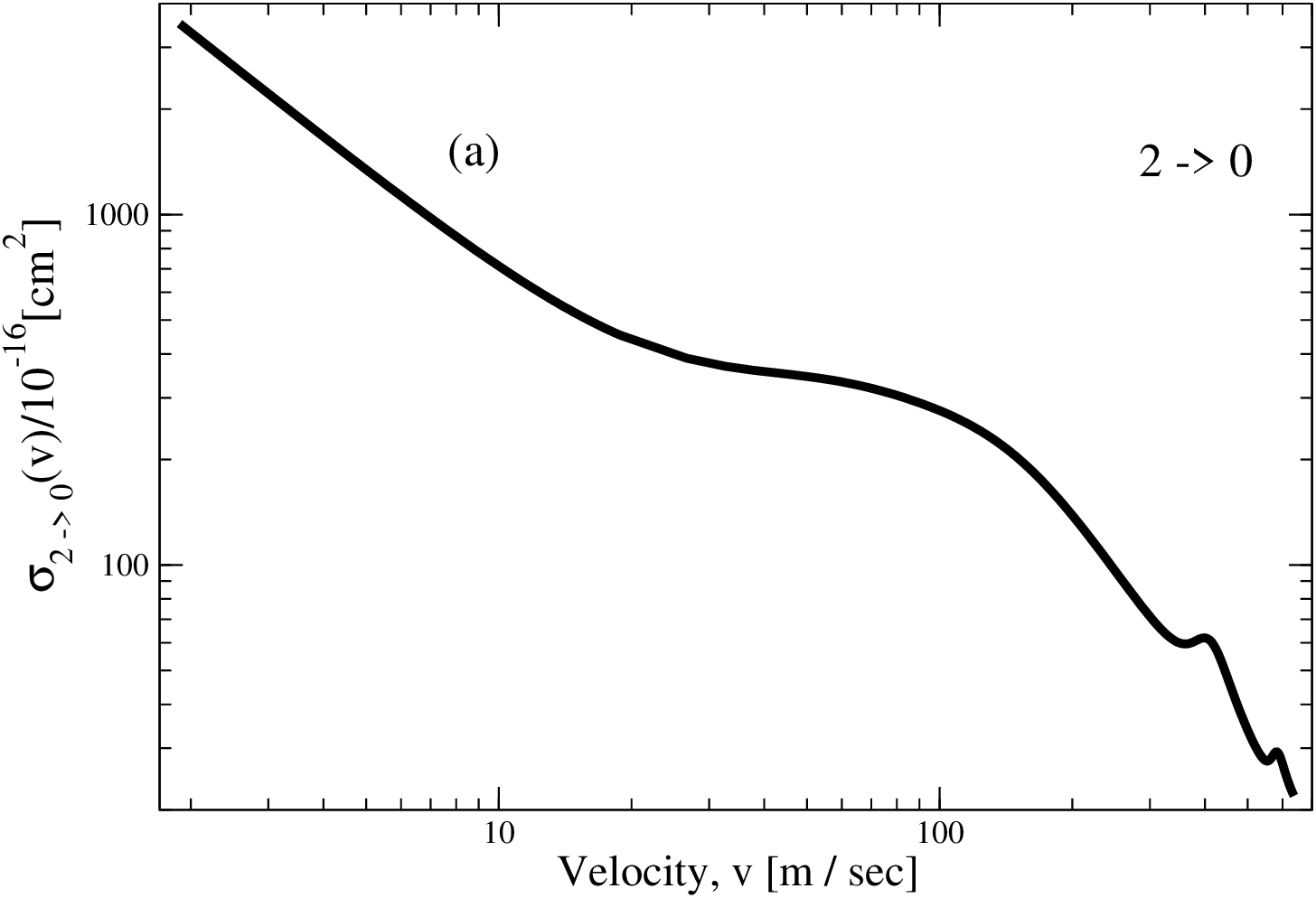}
\vspace{1cm}\\
\includegraphics*[scale=1.0,width=27pc,height=15pc]{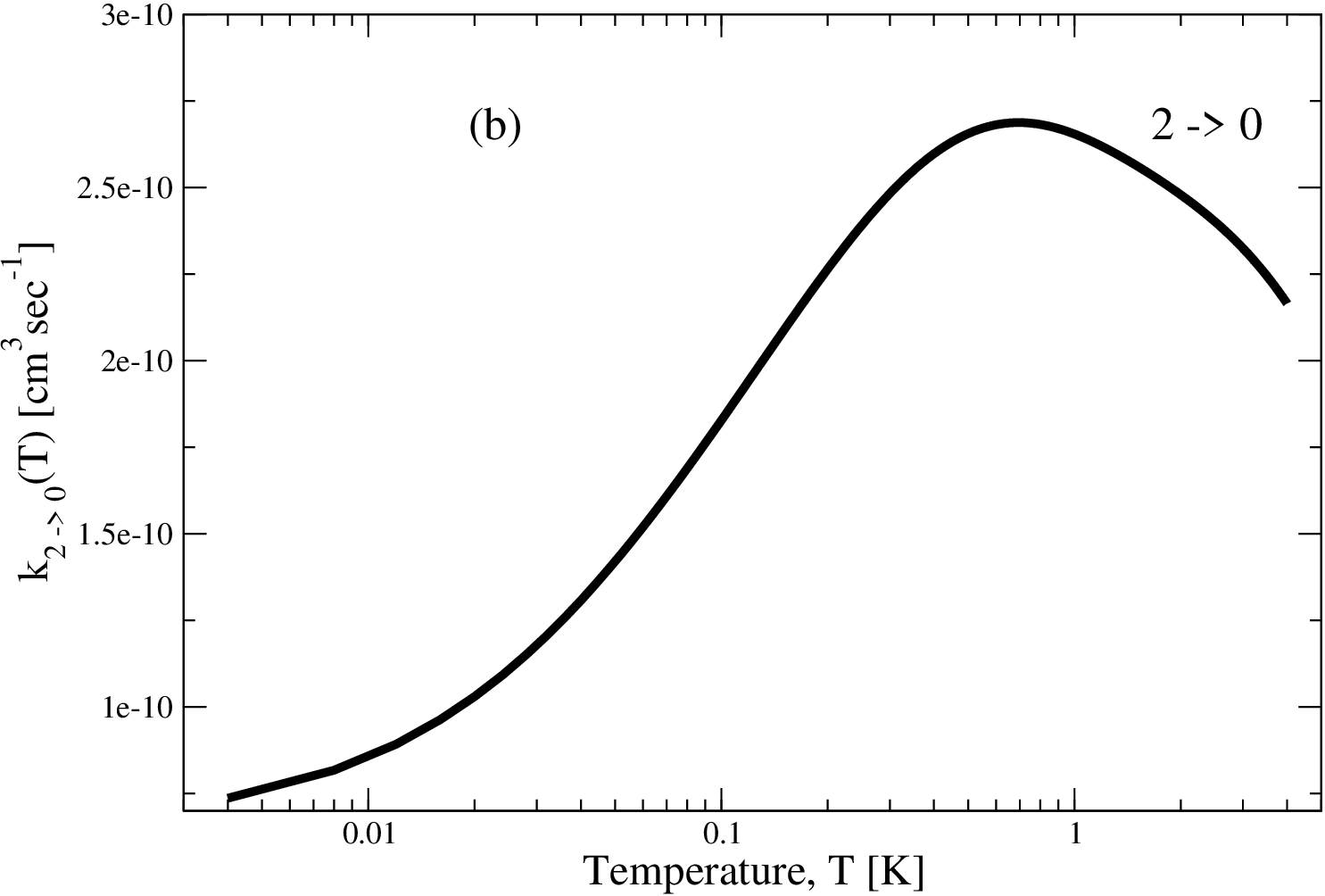}
\end{center}
\caption{ Upper plot (a): total state-resolved cross section
$\sigma_{2\rightarrow 0}(\mbox{v})$ vs. velocity $\mbox v$.
Lower plot (b): corresponding thermal rate
coefficients 
$k_{2\rightarrow 0}(\mbox{T})$ vs. temperature $\mbox T$
for the hydrogen molecule rotational relaxation process 
H$_2(j=2 \rightarrow j=0)$ 
in $\overline {\mbox{H}}$-H$_2$ collision.}
\label{fig5}
\end{figure}

\clearpage

\begin{figure}
\begin{center}
\includegraphics*[scale=1.0,width=27pc,height=15pc]{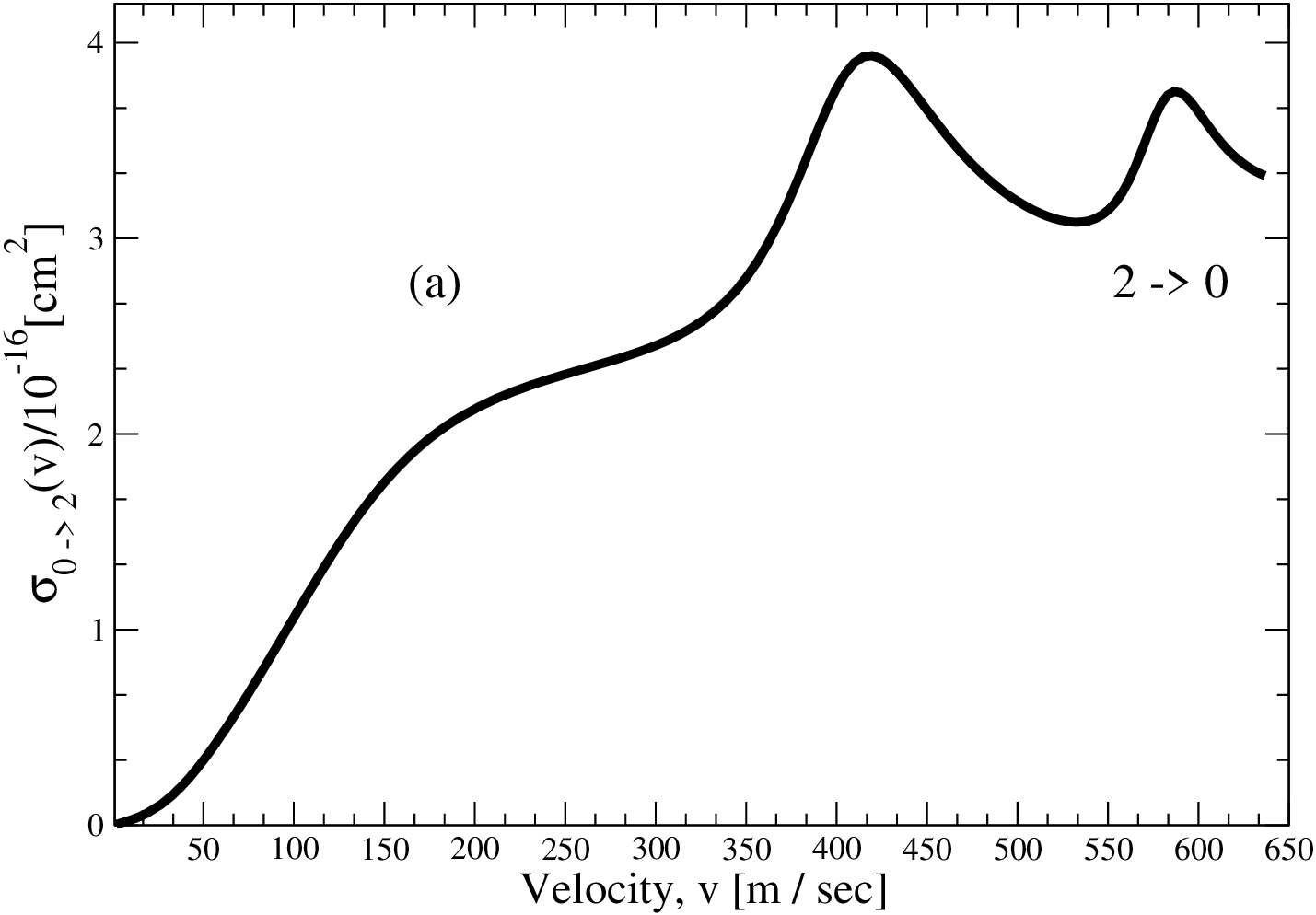}
\vspace{1cm}\\
\includegraphics*[scale=1.0,width=27pc,height=15pc]{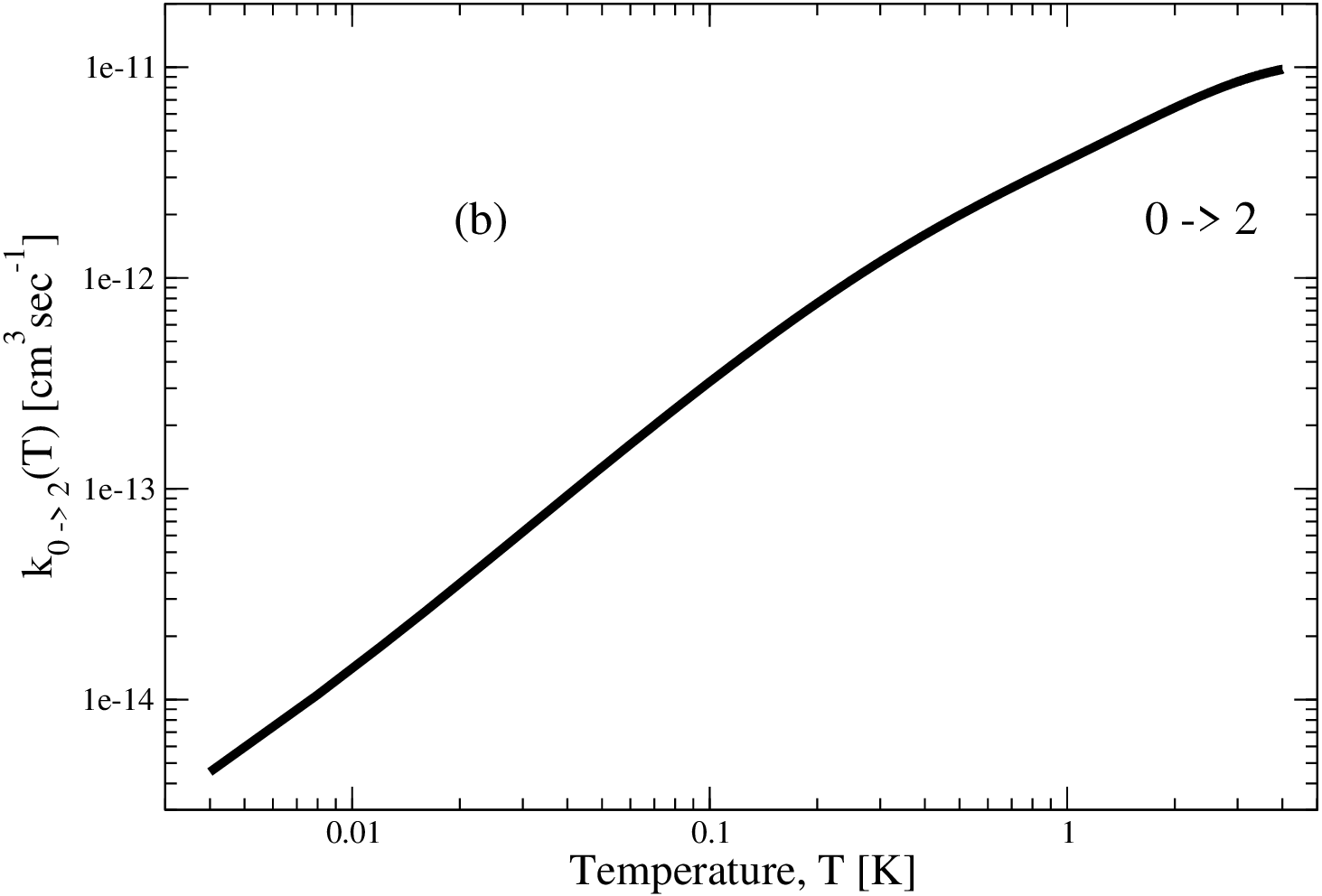}
\end{center}
\caption{ Upper plot (a):  total state-resolved cross section
$\sigma_{0\rightarrow 2}(\mbox{v})$ vs. velocity v. Lower plot (b):
corresponding thermal rate coefficients 
$k_{0\rightarrow 2}(\mbox{T})$
vs. temperature T
for the hydrogen molecule
rotational excitation  process H$_2(j=0 \rightarrow j=2)$
in 
$\overline {\mbox{H}}$-H$_2$
collision.}
\label{fig6}
\end{figure}

\clearpage

\begin{figure}
\begin{center}
\includegraphics*[scale=1.0,width=27pc,height=15pc]{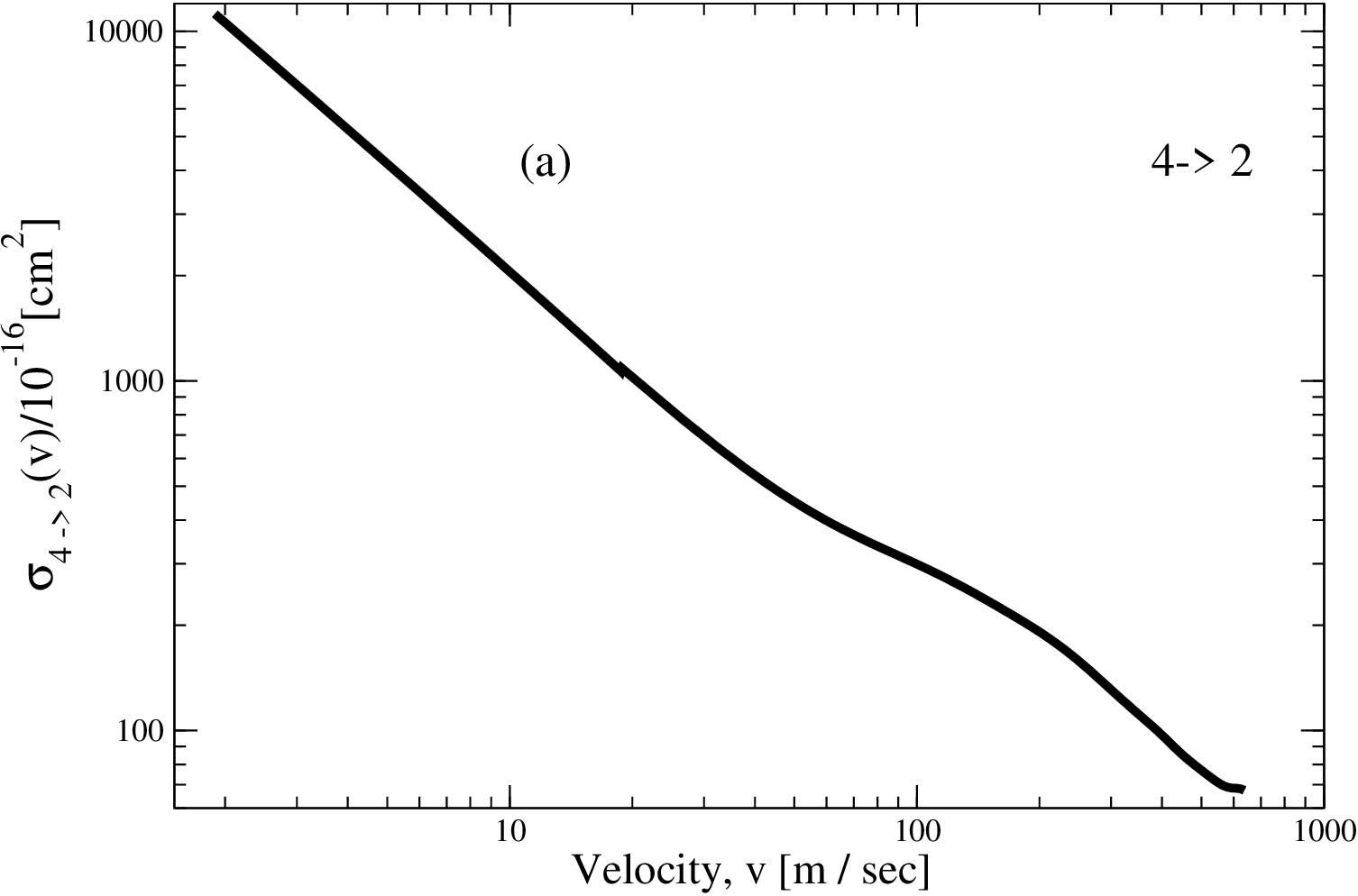}
\vspace{1cm}\\
\includegraphics*[scale=1.0,width=27pc,height=15pc]{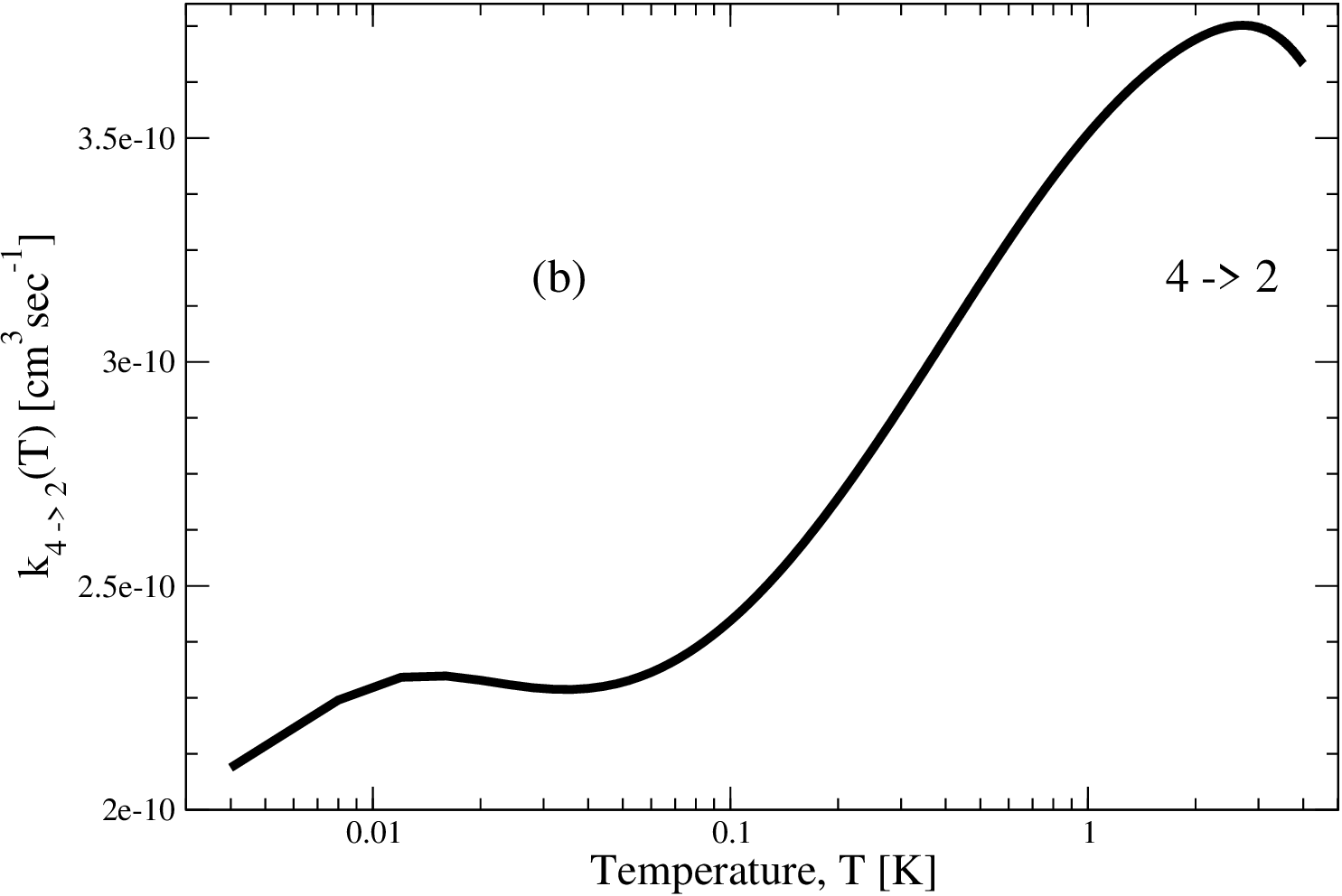}
\end{center}
\caption{ Upper plot (a): total state-resolved cross section
$\sigma_{4\rightarrow 2}(\mbox{v})$ vs. velocity $\mbox v$.
Lower plot (b): corresponding thermal rate coefficients
$k_{4\rightarrow 2}(\mbox{T})$ vs. temperature $\mbox T$
for the hydrogen molecule rotational relaxation process
H$_2(j=4 \rightarrow j=2)$
in $\overline {\mbox{H}}$-H$_2$ collision.}
\label{fig8}
\end{figure}

\clearpage

\begin{figure}
\begin{center}
\includegraphics*[scale=1.0,width=27pc,height=15pc]{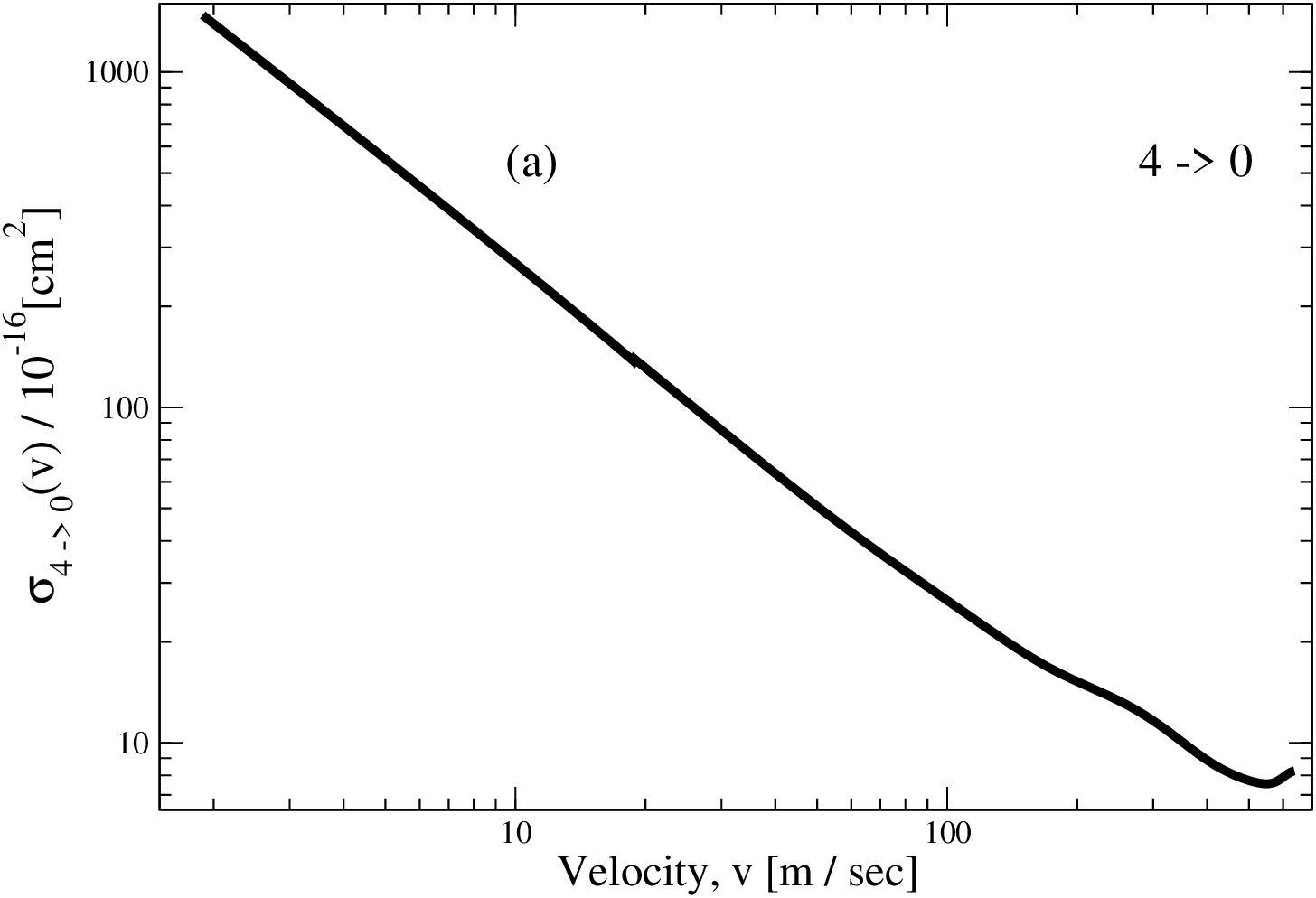}
\vspace{1cm}\\
\includegraphics*[scale=1.0,width=27pc,height=15pc]{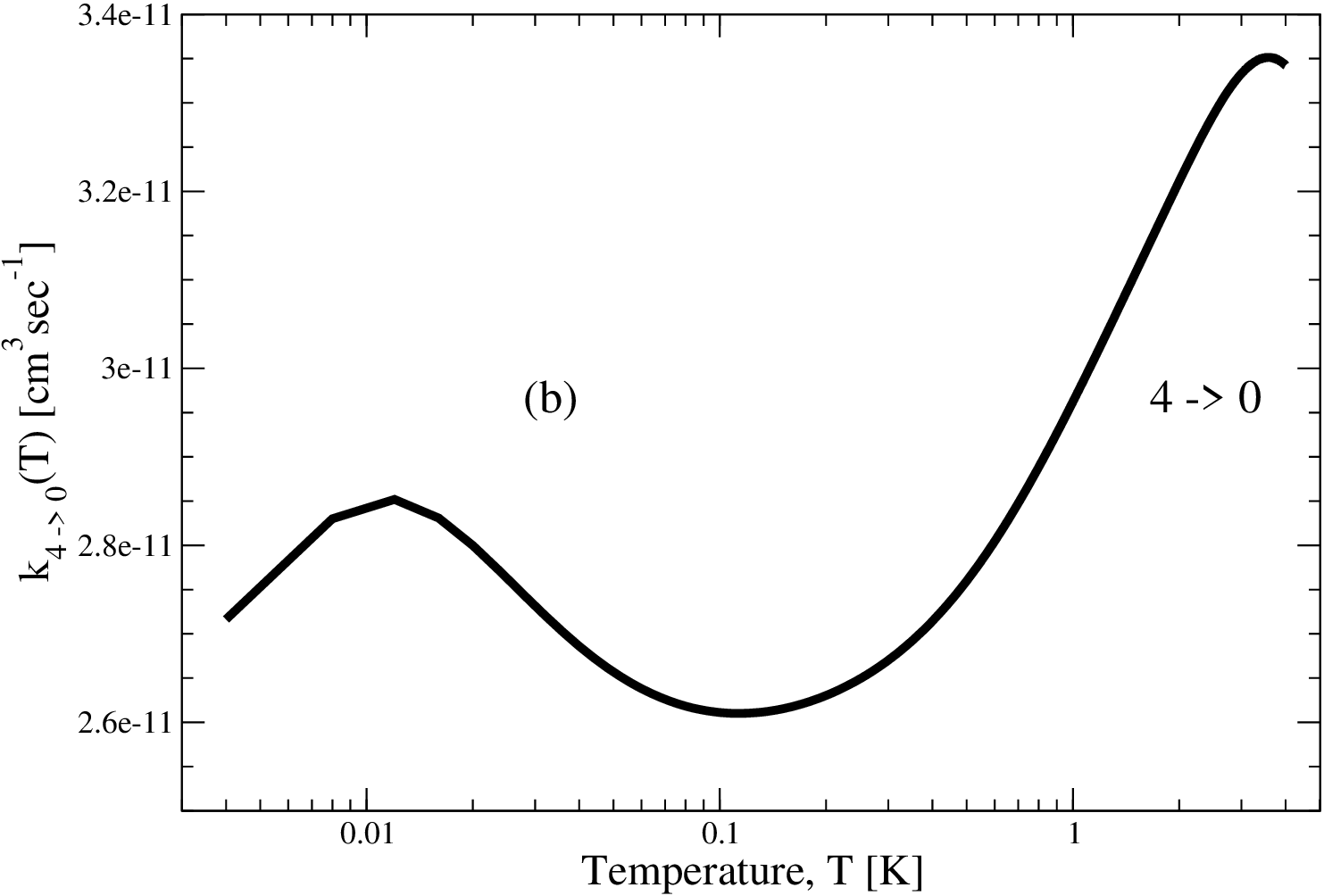}
\end{center}
\caption{ Upper plot (a): total state-resolved cross section
$\sigma_{4\rightarrow 0}(\mbox{v})$ vs. velocity $\mbox v$.
Lower plot (b): corresponding thermal rate
coefficients $k_{4\rightarrow 0}(\mbox{T})$ vs. temperature $\mbox T$
for the hydrogen molecule rotational relaxation process
H$_2(j=4 \rightarrow j=0)$
in $\overline {\mbox{H}}$-H$_2$ collision.}
\label{fig7}
\end{figure}


\begin{thebibliography}{99}
\bibitem{renat06}
R.A. Sultanov, D. Guster, Chem. Phys. 326, 641 (2006).
(Note that in Eq. (1)  of this work  
the momentum operators should be squared and the equation should look
as follows:
%
$
\left(\frac{P^2_{\vec R}}{2M_{12}}+\frac{L^2_{\hat
    r_{1}}}{2\mu_1r_1^2}+
\frac{L^2_{\hat r_{2}}}{2\mu_2r_2^2}+
V(\vec r_1,\vec r_2,\vec R) - E \right)
\Psi(\hat r_{1},\hat r_{2},\vec R)=0,
\label{eq:schred1}
$
as Eq. (2) in \cite{renat09}.
\end{thebibliography}
\end{document}